\documentclass[fleqn,usenatbib]{mnras}

\usepackage[T1]{fontenc}
\usepackage{ae,aecompl}

\usepackage{graphicx}	
\usepackage{amsmath}	
\usepackage{amssymb}	

\title[AARTFAAC transient survey]{The AARTFAAC 60\,MHz transients survey}

\author[M. J. Kuiack et al.]{
Mark Kuiack,$^{1}$\thanks{E-mail: m.j.kuiack@uva.nl}
Ralph A.M.J. Wijers$^{1}$
Aleksandar Shulevski,$^{1}$
Antonia Rowlinson,$^{1,2}$\newauthor
Folkert Huizinga,$^{1}$
Gijs Molenaar,$^{3}$
Peeyush Prasad,$^{1}$
\\
$^{1}$ Anton Pannekoek Institute, University of Amsterdam, Science Park 904, 1098 XH Amsterdam, The Netherlands \\
$^{3}$ Department of Physics and Electronics, Rhodes University, PO Box 94, Grahamstown, 6140, South Africa \\
$^{2}$ ASTRON, The Netherlands Institute for Radio Astronomy, Postbus 2, 7990 AA, Dwingeloo, The Netherlands \\
}

\date{Accepted XXX. Received YYY; in original form ZZZ}

\pubyear{2021}
\usepackage{newtxtext,newtxmath}
\begin{document}
\label{firstpage}
\pagerange{\pageref{firstpage}--\pageref{lastpage}}
\maketitle

\begin{abstract}
We report the experimental setup and overall results of the AARTFAAC wide-field radio survey, which consists of observing the sky within 50$^\circ$ of Zenith, with a
bandwidth of 3.2\,MHz, at a cadence of 1\,s, for 545\,h. This yielded nearly 4 million snapshots, two per second, of on average 4800 square degrees and a sensitivity of around 60\,Jy. We find two populations of transient events, one originating from PSR\,B0950$+$08 and one from strong ionospheric lensing events, as well as a single strong candidate for an extragalactic transient, with a peak flux density of $80\pm30$\,Jy and a dispersion measure of $73\pm3\,\mathrm{~pc~cm^{-3}}$, We also set a strong upper limit of 1.1 all-sky per day to the rate of any other populations of fast, bright transients. Lastly, we constrain some previously detected types of transient sources by comparing our detections and limits with other low-frequency radio transient surveys. 
\end{abstract}




\section{Introduction}

Discoveries of astronomical transients not infrequently occur by chance, during observations designed for other purposes, e.g., in case solar-flare gamma-ray outbursts  \citep{1959JGR....64..697P}, radio pulsars \citep{1968Natur.217..709H}, and gamma-ray bursts \citep{Klebesadel+1973}. Some, however, are the result of deliberately designed surveys, either to explore a new part of parameter space
\citep[e.g.,][]{2016MNRAS.456.2321S}, or to search for more examples of known type of transient, in recent history notably fast radio bursts \citep[e.g., CHIME \& ARTS,][]{2019Natur.566..230C,Oostrum+2020}. Others are set up with a specific goal, and then find or study a great wealth of other phenomena, such as the OGLE experiment \citep[see, .e.g.,]{1996IAUS..169...93P}. By now of course there are many such experiments at optical wavelengths \citep{2004SPIE.5489...11K, 2009PASP..121.1334R, 2019PASP..131a8002B}.

The discovery of as-of-yet unobserved phenomena in a new part of parameter space
has been a key science goal of the Low-Frequency Array \mbox{\citep[LOFAR;][]{2013A&A...556A...2V}} since its inception \citep{2006smqw.confE.104F}.
AARTFAAC, the Amsterdam-ASTRON Radio Transient Facility and Analysis Center, is a project developed to utilize LOFAR's central stations to expand its transients discovery space to even wider fields and faster cadences, but at relatively poor angular resolution and sensitivity: it searches for the rarest and brightest  transient and variable phenomena on seconds time scale. 
As a dedicated transient detector, its goal is to maximize the useful information recorded, near-real time, and generate alerts for broadcast to the multi-wavelength transient community. 

Beyond discovering entirely new classes of transient events, our blind survey allows us to constrain surface density-brightness limits on the low-frequency counterparts to known transient and variable sources.
Even upper limits to the prompt emission from mergers of binary neutron stars or neutron-star black-hole binaries could aid in determining the neutron star equation of state; for gamma-ray bursts (GRBs) and magnetar outbursts, the  host environment and emission physics could be constrained \citep[e.g.,][and references therein]{Gaensler+2005, GranotHorst2014,Burlon+2015}; for fast radio bursts (FRBs) the rates and/or the low-frequency spectrum and emission mechanism might be constrained  \citep{2017MNRAS.465.2286R}.

Realizing the goal of a near-real time transient detection requires the development of algorithms that calibrate and image the radio data very rapidly and then automatically filter an enormous number of spurious candidates from the large number of potential transients
found in the resulting image stream. 
This in turn requires thorough characterisation of the instrument and data. 
Therefore, while the primary purpose of our survey is to detect bright, low-frequency,  transient events that evolve on timescales of seconds, our secondary purpose is the development of a detection pipeline which can monitor our data stream and produce a reasonable number of candidates for closer inspection. 

We performed a blind survey at 60\,MHz for transient events with timescales down to one second. 
In doing so we developed a novel method for analysing light curves to detect any potentially interesting events, while reducing detections of specious sources, e.g., noise, interference, satellites, and airplanes.
With that, we observed various phenomena, from extreme-fluence pulsar outbursts \citep{2020MNRAS.497..846K}, to isolated scintels magnifying persistent sources and scintillating structures lasting many hours \citep{2021MNRAS.tmp.1149K}. In this paper we present the overall data analysis strategy of the survey and limits to the surface densities of any as yet undiscovered phenomena, and compare these results with other recent low-frequency transient surveys. As well as reporting the discovery of a potential cosmic transient.

In Sect. \ref{sec:obs} we briefly describe the instrument and data products analysed, and the set of observations gathered. Next, in Sect. \ref{sec:methods} we outline the method we have  developed to analyse our source databases, and detect potentially astrophysically  interesting events. Our results are presented in Sect. \ref{sec:results}, followed by a discussion of our findings in the context of other recent surveys in Sect. \ref{sec:discussion}. Lastly, we conclude ins Sect. \ref{sec:conclusion}.

\section{Instrument and Observation}
\label{sec:obs}

\subsection{Instrument}

\begin{figure*}
\includegraphics[width=0.8\textwidth]{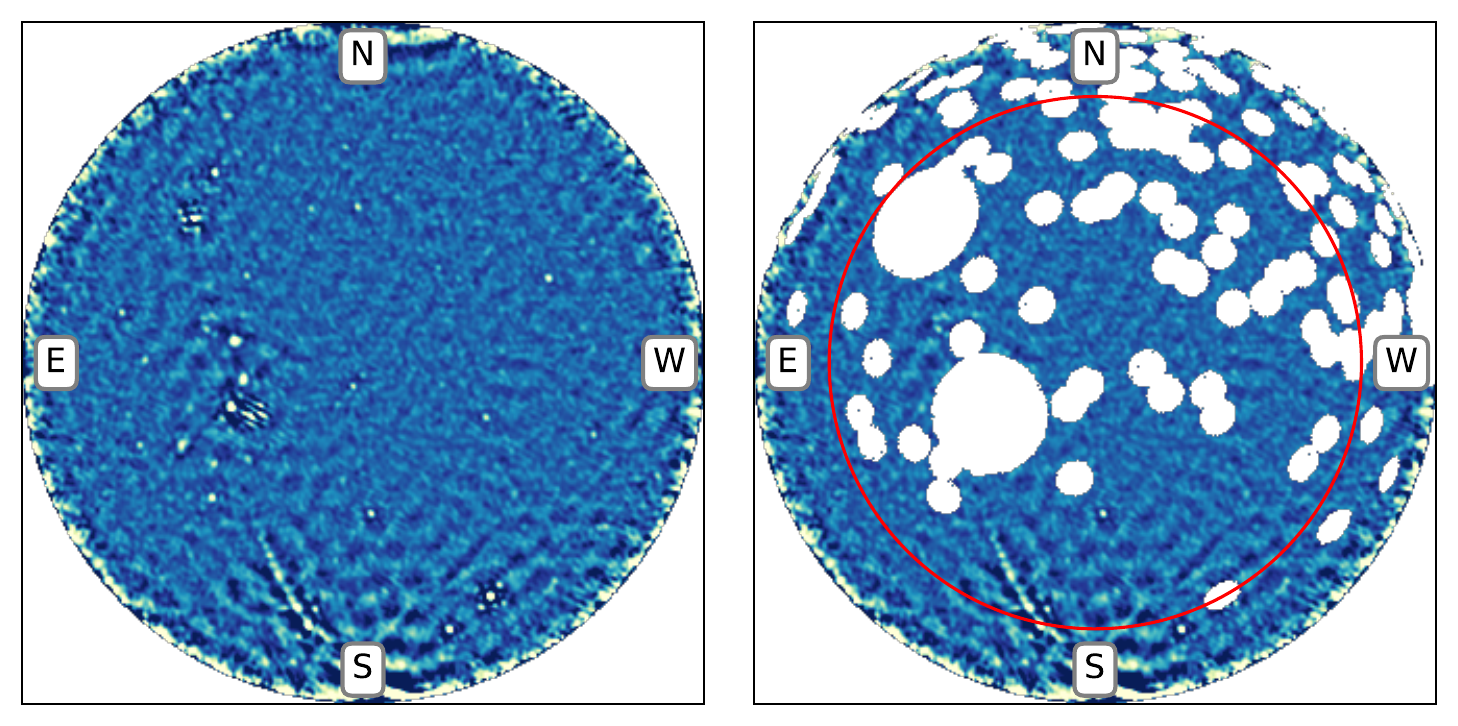}
\caption{Left: An example single AARTFAAC snapshot image. Right: The remaining sky area surveyed after regions around the AARTFAAC catalogue sources and A-team have been excluded. The red circle indicates the extent of the detection region, 50$^\circ$ from Zenith.
The outer edge of the image is at elevation 4$^\circ$, in order to cut off the worst noise close to the horizon. After excluding $3^{\circ}$ around the AARTFAAC catalogue sources, and $10^{\circ}$ Cassiopeia A and Cygnus A, 70\% of the area remains search-able. In both images, halfway South from Zenith, a magnified source, normally below our detection threshold, is visible.}
\label{fig:sky_map}
\end{figure*}

AARTFAAC  is a multi-purpose, wide-field, fast cadence imaging, low-frequency radio project \mbox{\citep{2014A&A...568A..48P,2016JAI.....541008P}}. 
It operates as a parallel back end, utilizing the low-band antennae (LBA) simultaneously during other regularly scheduled LOFAR observations. Since AARTFAAC can only operate during LBA observations and some otherwise unscheduled (`filler') time, the observation times and duration are somewhat random, and a single observation can last from under 1\,h to 24\,h.
Signal is taken directly from 288 dipoles in the outer rings of the central six LOFAR stations, providing a maximum baseline of 300\,m and yielding a spatial resolution of 1$^\circ$ at 60\,MHz.
The dipole response pattern, together with our correlating, calibration and imaging pipeline produce Zenith-pointing snapshots with a sensitivity to the full sky, decreasing from best at Zenith to very poor on the horizon \citep{2014A&A...568A..48P,2016JAI.....541008P}.

Complex gain calibration, for both direction dependent and direction independent effects, is performed in real-time using a multi-source self calibration algorithm \citep{2009ITSP...57.3512W}. Calibration is done on each time step and frequency subband independently; to increase throughput converged solutions from previous time steps are used as starting points for the solution at the next time step. The sky model used is composed of the 5 brightest sources in the radio sky, Cassiopeia A, Cygnus A, Taurus A, Virgo A, and the Sun. Each is modelled as a single Gaussian, an accurate enough representation given AARTFAAC resolution.
In order to reveal the fainter sources the bright calibrators are then subtracted. A full detailed description of the calibration method is given by \cite{2014A&A...568A..48P}.

Calibrated visibilities are then streamed live to disk and to an imaging pipeline which performs an FFT on the gridded visibilities. Additionally, sensitivity to the large scale Galactic emission is reduced by removing baselines shorter than $10\lambda$ (where $\lambda = 5$\,m at 60\,MHz). 
Lastly, a primary beam correction is applied, which flattens the antenna response across the field of view \citep{10.1093/mnras/sty2810}. 
An example of a snapshot image is given in the left-hand panel of Fig. \ref{fig:sky_map}. This yields a typical background noise of 7.5\,Jy/beam  at Zenith for images generated by integrating 8 consecutive sub-bands. This is somewhat less than the theoretical confusion noise limit of 10.4\,Jy/beam \citep[Eqn. 6;][]{2013A&A...556A...2V}.  The full pipeline introduces a total latency of around 1.5 seconds from the received signal to the output image.

\subsection{Observations}

In this work we present the analysis of 545.25 hours of data 
(after rejection of bad images) collected between August 2016 and September 2019. 
The distribution of observations is shown in Table~\ref{tab:Obslist}.

\begin{table}
\begin{center}
\caption{The time distribution and duration of the observations which were analysed for this work.}
\begin{tabular}{c c c c}
\hline \hline
Month  & Hours observed & Month & Hours observed  \\  
\hline
2016-08 & 2.5 & 2018-10 & 70.9 \\
2016-09 & 4.9 & 2018-11 & 55.5 \\
2016-11 & 0.6 & 2018-12 & 15.4 \\
2017-02 & 31.6 & 2019-01 & 111.0 \\
2018-03 & 26.8 & 2019-02 & 34.1 \\
2018-04 & 41.9 & 2019-03 & 31.5 \\
2018-07 & 24.2 & 2019-05 & 63.4 \\
2018-09 & 24.7 & 2019-09 & 6.2 \\
\hline \hline
\label{tab:Obslist}
\end{tabular}
\end{center}
\end{table}

The data were recorded at 1\,s time resolution, with 16 sub-bands of width 195.3\,kHz, arranged in two blocks of 8 consecutive sub-bands, from 57.5 -- 59.1 and 61.0 -- 62.6 MHz. 
Each frequency block was then summed to create two images per second.
The image flux scaling is calculated independently per image by comparing the sources, extracted using PySE \citep{2018A&C....23...92C}, to the AARTFAAC catalogued values, using the method described in  \cite{10.1093/mnras/sty2810}. Prior to flux scaling, poorly calibrated images are rejected if the rms.\ of the image data is greater than 1000 arbitrary flux density units ( about 3400\,Jy/beam). (Typical rms.\ values for well calibrated images, including sources and pixels down to the horizons where the noise drastically increases, are around 50 (170\,Jy/beam); this rejects a widely varying fraction of the data, from 60\% in the poorest conditions when calibration often fails, to only 0.9\% in the best, see \cite{10.1093/mnras/sty2810}.) 545.25\,h is thus the  total amount of data used for transient detection, excluding all poorly calibrated images and time steps where only one of the two frequency intervals was successfully imaged.

For detecting transients in astronomical images we use the LOFAR Transient Pipeline v3.1.1\footnote{\url{https://github.com/transientskp/tkp}} \citep[TraP;][and references therein]{2015A&C....11...25S}. 
TraP processes the images in sequence, extracting sources and then combining all detections of the same source into one, with a light curve, by making associations with a database of sources previously detected during the observation.
Due to the computational limits of searching a large database, we create a new independent database for each observation.
We use a detection threshold of $5\sigma$, with a detection radius of 400 pixels from the image centre, which corresponds to 50$^\circ$ from Zenith (where the primary beam gain has dropped to 0.53 the gain at zenith) and an area of $7368~\mathrm{deg}^{2}$.
The outer radius is chosen to maximize the searchable area, by setting the edge where the sensitivity is still reasonable and near-horizon effects like RFI and low antenna gain are still small.  

\section{Methods}
\label{sec:methods}

Once fast imaging, calibration, and source detection have been achieved,
the primary remaining challenge in transient detection  with a sensitive wide-field instrument, such as AARTFAAC, is the great number of transient and variable sources that are not astrophysically interesting. 
Transient sources include terrestrial radio frequency interference (RFI), detected from the horizon, or reflected from ionized meteor trails, airplanes, or originating from satellites. 
Additionally, propagation of distant stable signals through the ionosphere and plasmasphere can be strongly modulated to appear transient, or variable \citep{2021MNRAS.tmp.1149K}. 
Excluding these different types of false transient candidate requires a multi-staged filtering process. 

The first step in our filtering process is setting the detection threshold sufficiently high to exclude the events occurring in the tail of the background noise distribution. 
Our initial detection threshold per source is $5\sigma$, which resulted in $3.2\times10^5$ detections over the 545.25 hours (where one detection means one light curve added to the catalogue, i.e., a unique sky location
with at least one, but possibly very many  detections during one observing interval).
We exploit the two images we make per time-step, by requiring a source to be detected in both images, with at least one $8\sigma$ detection in either frequency. 
This filters a large amount of RFI, which tends to be narrow band, and random noise occurring  independently in either image. This filters out roughly 85\% of detected transient candidates. 

Secondly, a large number of spurious events, 10\% of all detections, are filtered out by excluding regions around known bright, persistent sources. 
The AARTFAAC catalogue \citep{10.1093/mnras/sty2810} is slightly deeper, so therefore many sources are at the detection threshold of the transient survey. 
These sources will scintillate above the threshold and create false positives. 
During times of extreme scintillation the sidelobes of the brightest sources will also be above the threshold. 
We therefore exclude 3$^\circ$ around the AARTFAAC catalogue sources, and 10$^\circ$ around the A-team sources. 
These regions err on the cautious side, because the reduction in false positives is worth the sacrifice of surveyed area. 
The region of surveyed sky in an example image is presented in the righthand panel of Fig.~\ref{fig:sky_map}. 
The remaining sky area is typically between 60 and 70\% of the sky, depending on the specific local sidereal time (LST). The average fraction of the detection region surveyed is 0.65, resulting in a mean survey area per image of $\rm 4789~deg^{2}$.

Due to the exclusion regions the list of candidates is limited to regions of sky where there are typically no bright sources.
All source association is done in celestial coordinates, without accounting for possible
proper motion.
Therefore, objects moving with respect to the celestial background, such as airplanes and satellites moving through our field of view, generate a multiplicity of sources across their trajectory, each appearing in the catalogue for that observation as a separate light curve of a false transient. Sometimes the motion is slower than one PSF width per image, and the object may appear as a set of short light curves of a smaller number of unique sources. 
In any case, seen over a significant time interval these sources present as streaks across the image, which are
obvious artefacts that can easily be recognized and removed, or ignored, with a variety of machine learning techniques, given also that true astrophysical transients are rare (Section~\ref{sec:discussion}). We found that excluding new source detections that occur within a space-time distance of 2.5$^\circ$ angular separation and 500\,s was adequate to reject moving sources; these parameters were tuned using a sample dataset which contained a number of transiting unidentified flying objects. In Table~\ref{tab:Filterlist}, we give the number of sources left in the sample after application of each of these filtering criteria.

Finally, the 9061 sources remaining were manually inspected and then those not rejected as spurious were analysed automatically using the light-curve peak detection and analysis technique described in \cite{2021MNRAS.tmp.1149K}. Still quite a few of these were found to be trivial, such as more sidelobes of bright sources and persistent sources below our detection threshold that occasionally brighten enough to cause a detection. Therefore, there is more still that can be done to reduce the number of false transients in an automated real-time pipeline. The astrophysically relevant transients are discussed further in Section~\ref{sec:respositive}.

\begin{table}
\begin{center}
\caption{The total number of sources, defined as a $5\sigma$ detection of flux density originating from a unique celestial coordinate, per observation, which were left after applying each successive filtering criterion.}
\begin{tabular}{l r}
\hline \hline
 Filter step & Source count  \\  
\hline
Initial TraP output  & 322,571 \\
rejected due to $<8\sigma$ in one band & 271,188  \\
rejected due to catalogue match &  35,196 \\
rejected moving sources & 6,478  \\
Remaining & 9,709  \\
\hline \hline
\label{tab:Filterlist}
\end{tabular}
\end{center}
\end{table}

\section{Results}
\label{sec:results}
\subsection{Flux and position uncertainty}

We investigate the validity of our flux density calibration by comparing all flux density measurements of the AARTFAAC catalogue sources made during the survey.
We find the mean flux ratio, $f_{\mathrm{survey}}/f_{\mathrm{cat}} = 1.06 \pm 0.25$. 
The 25\% flux density uncertainty, predominantly due to ionospheric variability, is in line with our previous results \citep{10.1093/mnras/sty2810}.

The 300\,m maximum baseline results in an angular resolution of 1$^\circ$. However, the final position uncertainty is also affected by the apparent random movement of sources due to scintillation, and systematic errors in the calibration and imaging pipeline. 
We find that apparent motion due to scintillation is not a significant factor, with the typical standard deviation of both RA and DEC being around $0.1^{\circ}$, well below the  PSF width.

\begin{figure}
\includegraphics[width=\columnwidth]{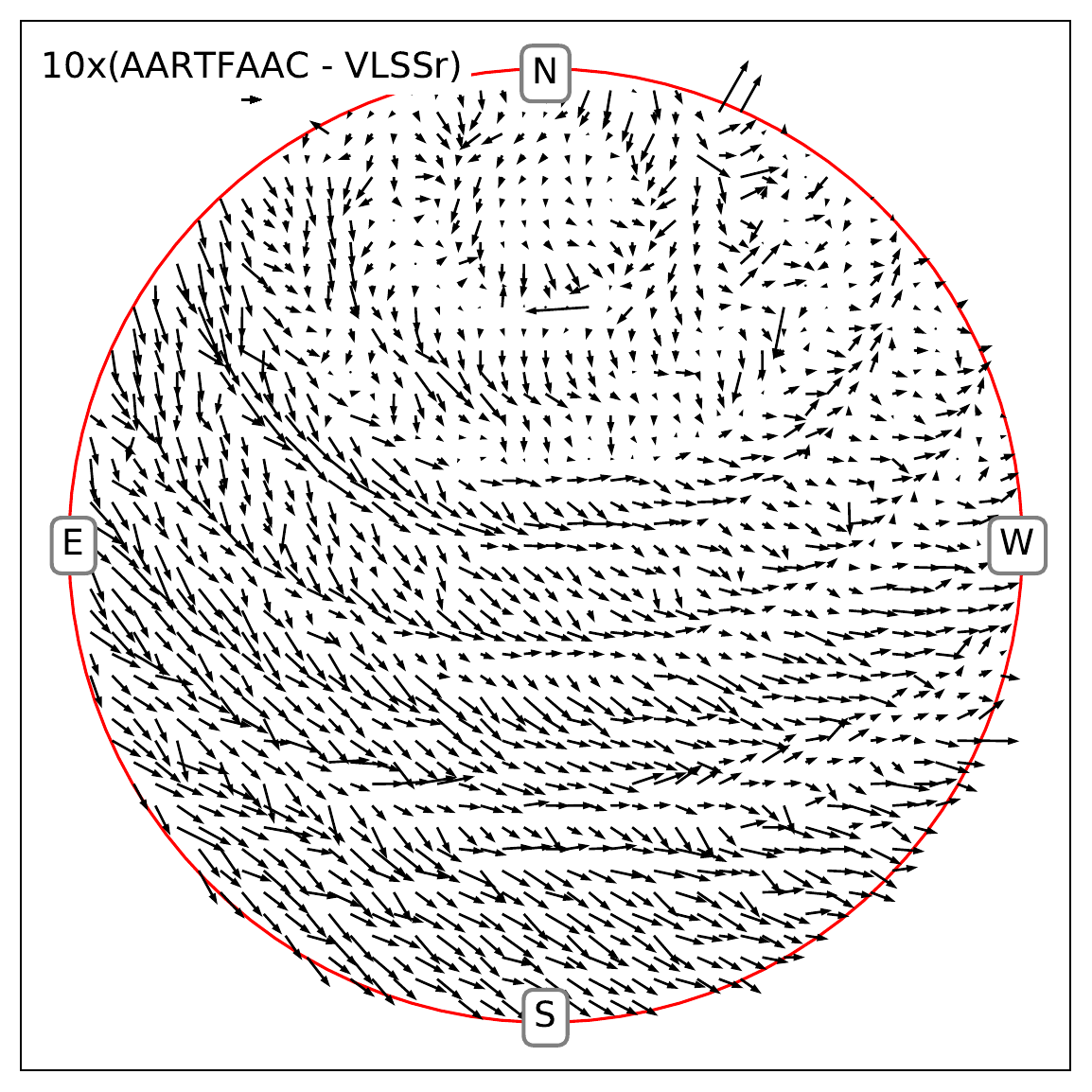}
\caption{The position difference between the AARTFAAC WCS and VLSSr sources. The arrow lengths are $10\times$ the real sky distance. The mean position offset is $0.25^{\circ}$.The red circle is again the edge of the search area at $Z=50^\circ$.}
\label{fig:posoffset}
\end{figure}

We also, however, measured a significant, systematic offset in our image WCS by comparing the measured positions of bright sources in the AARTFAAC images to the VLSSr catalogue. 
The direction of the offset is not constant around the image but varies depending on the source's location on the sky. 
In Fig.~\ref{fig:posoffset} we show how the position difference changes as a function of horizontal coordinates. 
The interpolated grid of position differences was generated with the full AARTFAAC survey. This indicates that it is an error in the calibration or imaging pipeline. 
With a mean distance of $0.25^{\circ}$, the correct coordinate is still within an AARTFAAC beam width and thus of no consequence to our source identification.
The error was known and corrected ad-hoc in our recent analysis, such as the association of extreme-fluence pulses with B\,0950+08 \citep{2020MNRAS.497..846K}.

\subsection{Sky area, coverage and Sensitivity}

\begin{figure*}
\includegraphics[width=\textwidth]{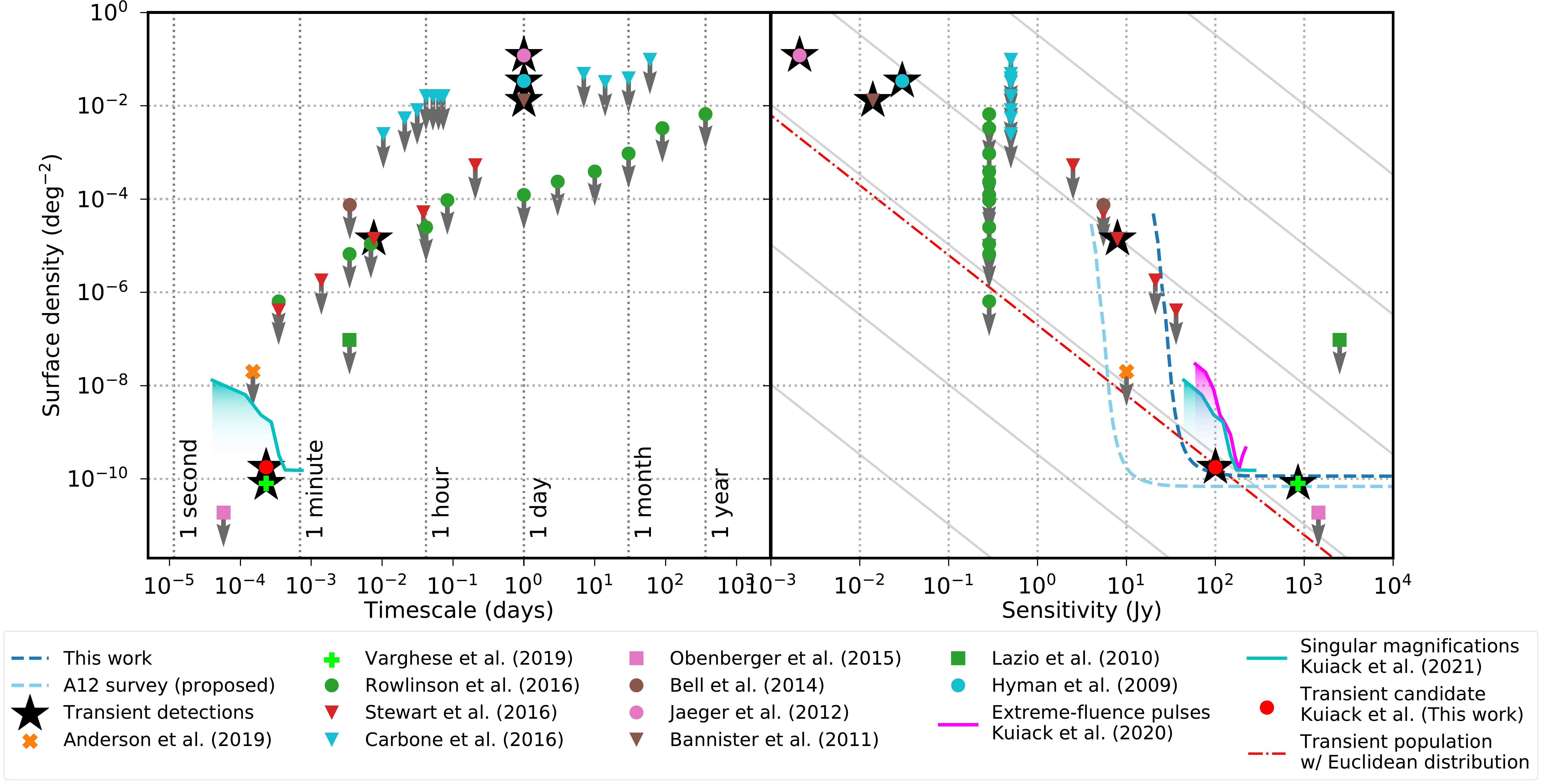}
\caption{ We compare the surface density to which we have probed for transients, that is the $8\sigma $ detection limit, as a function of timescale (left) and sensitivity (right), to a multitude of other recent surveys. Arrows denote upper limits placed by the survey, black stars mark reported transient detections. The diagonal lines in the right-hand panel have a power-law index of $-3/2$,  expected of a population of standard candles uniformly distributed in Euclidean space, the red dash-dot line is scaled to the surface density of the single potential extragalactic transient  candidate discovered during this survey, and the grey lines scaled to the transient detected by Stewart et al. (2016), in steps of $10^{3n}$, for integer $n$.}
\label{fig:sensitivity}
\end{figure*}

Using the assumption that the rate of some unknown transient population would follow a Poisson distribution, the probability of detecting $n$ transients during a survey,
\begin{equation}
    P(n) = \frac{\lambda^n}{n!} e^{-\lambda}, 
\end{equation}
where $\lambda = \rho (N_{\mathrm{epochs}} -1)\Omega_{\mathrm{FOV}} $ is the expected number of transients, with  surface density $\rho$, observable at any given instant. The number of independent epochs, one second time steps with two good images,  in the AARTFAAC survey is $N_{\mathrm{epochs}} = 1,962,910$,  with an average usable search area of  $\Omega_{\mathrm{FOV}} = 4789 ~\mathrm{deg}^2$. 
With a non-detection, we can calculate the surface density to a 95\% confidence level by setting $P(0) = 0.05$ and solving for $\rho$,
which gives $\rho = 3.1 \times 10^{-10} ~\mathrm{deg}^{-2}$ (which is equivalent to 1.1 per day all-sky).

However, this assumes that the given transient is detectable in every region across every image, which neglects the fact that for a wide-field instrument such as AARTFAAC, the transient sensitivity can vary across the field of view in a single image, and across images according to local sidereal and universal time. We can refine this by summing the total surveyed area as a function of sensitivity limit, allowing us to plot a continuous distribution that describes the surface density surveyed at each sensitivity \citep{2016MNRAS.459.3161C}.
In Fig.~\ref{fig:sensitivity} we show this curve for the AARTFAAC survey (dark blue dashed curve), compared to many recent transient surveys across a wide range of frequencies and timescales.

The survey sensitivity coverage is calculated from the parameters stored in the database. 
Only information measured and calculated during each source extraction is stored, to reduce the data volume. 
Since the RMS noise and thus the sensitivity are slowly varying with position within the detection region, the survey database does not store a noise map, but only the flux and detection signal-to-noise of each source. We then of course know the RMS noise at each source location as the ratio of those numbers and can use those values to recreate a good noise map by interpolation. As
a boundary value for the noise at the edge of the detection region for use in this interpolation
we use the average of $\rm RMS_{\mathrm{max}}$ and $\rm RMS_{\mathrm{qc}}$, which are the maximum RMS in the image and RMS in the region 20$^\circ$ around Zenith.  
Both values are calculated and stored by TraP during image processing.  
This approximation has been empirically verified by  calculating the sensitivity area distribution directly from a set of 100 images across a six hour observation. 

\subsection{Positive detections}
\label{sec:respositive}

Our survey revealed three populations of transient events, observable due to our fast imaging cadence.

\subsubsection{Pulsar giant pulses}

The first are extreme pulses from the pulsar B0950+08 \citep{2020MNRAS.497..846K}. In the right-hand panel of Fig. \ref{fig:sensitivity} we show the surface density of these pulses detectable as a function of the sensitivity limit (magenta line). 
The distribution is the number of pulses of a given brightness, divided by the total sky area surveyed to that detection threshold. Since each pulse was detected blindly without prior assumption of a location, this fairly represents the surface density of such pulses, regarded as independent
events in a population. For the astrophysical implications we must of course
account for the fact that they all come from the same source, see \citep{2020MNRAS.497..846K}.

\subsubsection{Isolated scintels}

The second population of transients observed, indicated by the blue line in Fig. \ref{fig:sensitivity}, do appear to be independent events. These are the spontaneous, singular magnifications of background sources, likely due to ionized plasma within the ionosphere of the Earth \citep{2021MNRAS.tmp.1149K}. 
The curve illustrates the brightness distribution of these `scintels'.
They are time resolved, with widths typically ranging 10--100\,s, and therefore we also plot their rate distribution as a function of timescale in the left panel of Fig. \ref{fig:sensitivity}. Transient events of this class are also characterized by a time-frequency evolution that is inconsistent with an extragalactic dispersion delay.

\subsubsection{Potential cosmic transients}

\begin{figure}
\includegraphics[width=\columnwidth]{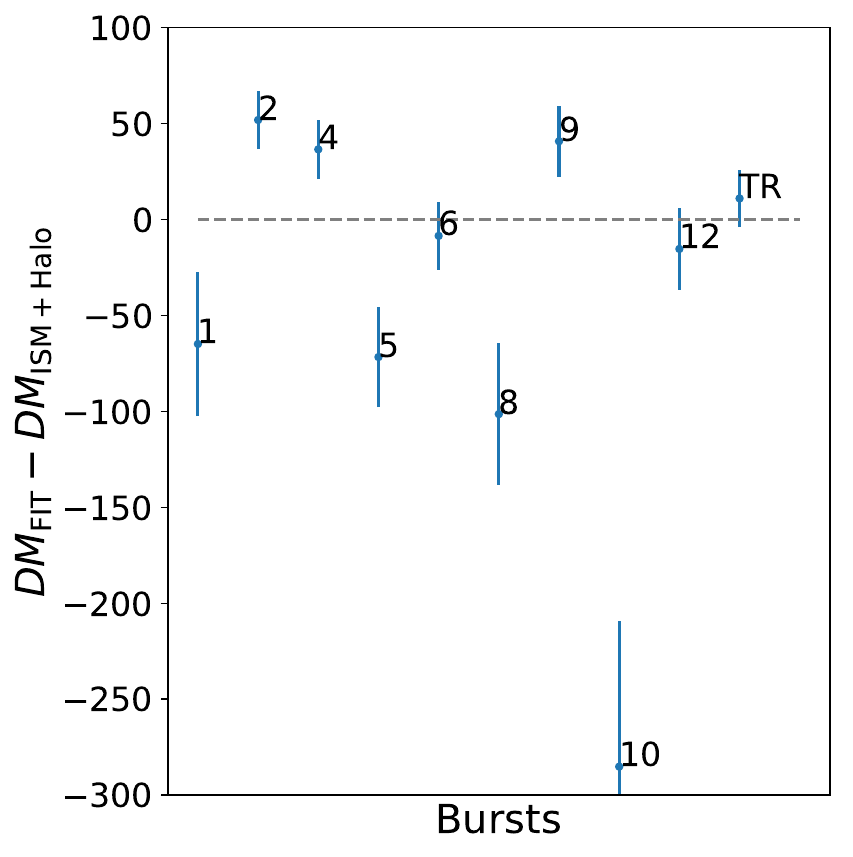}
\caption{The inferred excess DM beyond the contribution from the Galactic ISM and halo. Sources below the zero line can be rejected as extragalactic transients, whereas those above could either be extragalactic or might embedded in regions where the local ISM density is higher than predicted by the NE2001 model. We use a 20\% uncertainty in the model DM, which is much larger than our range.}
\label{fig:DMcompare}
\end{figure}

Lastly, we observed a sample of transient events that do exhibit a time frequency evolution across the 4.2\,MHz we observe, that could be consistent with dispersion delay due to propagation through the interstellar medium (ISM) and intergalactic medium (IGM). These candidates are summarized in Table \ref{tab:transients}. The candidates whose inferred dispersion measure (DM) is greater than the sum of the NE2001 ISM model \citep{2002astro.ph..7156C} and Galactic halo model \citep{2020ApJ...888..105Y} are shown above the dashed grey line in Fig. \ref{fig:DMcompare}. These sources are more interesting to us because they can be more easily differentiated from the non-dispersed isolated scintels, singular magnifications due to near-Earth plasma, which typically have shorter (positive and negative) inter-frequency delays \citep[Illustrated in fig. 10 of ][]{2021MNRAS.tmp.1149K}.

\begin{figure}
\includegraphics[width=\columnwidth]{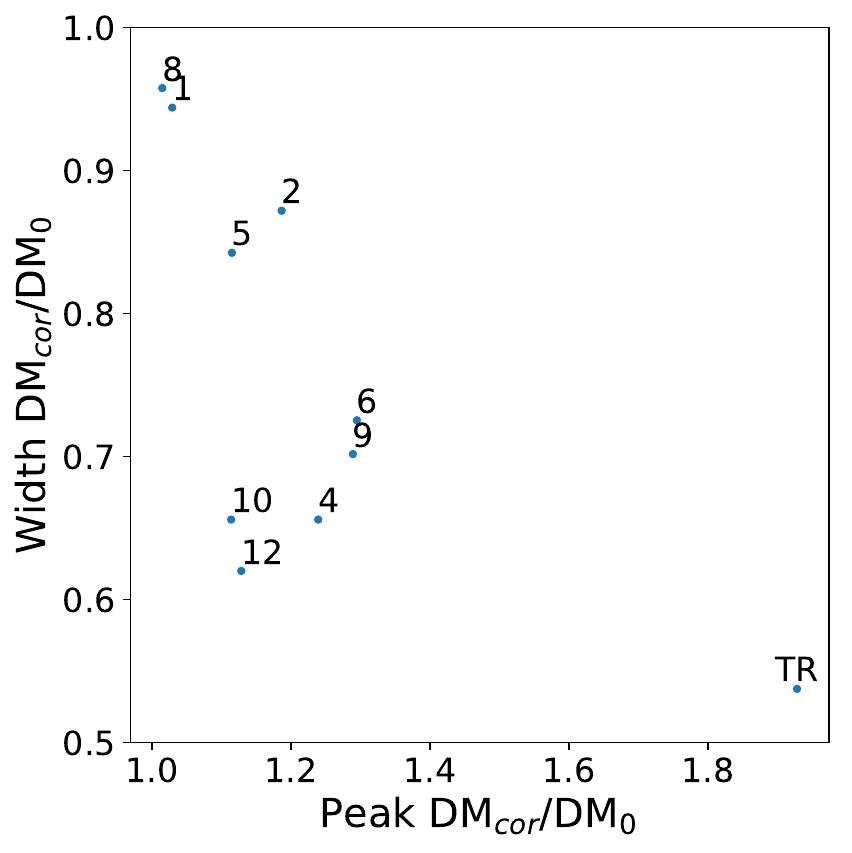}
\caption{Here the effect of DM correction on each candidate is illustrated by plotting the ratio of the corrected to uncorrected burst width against the corrected to uncorrected burst peak height. Uncorrected width and peak are measured by integrated our 16 frequency bands without any frequency delay, i.e., $\rm DM =0$, the $\rm DM_{\rm corr}$ corresponds to the best fit DM for each candidate.  Intrinsically narrower busts with higher dispersion appear further down and to the right where our strongest candidate, labelled TR, is. }
\label{fig:DMcorrected}
\end{figure}

We further differentiate this sample by comparing the relative effect that correcting the inferred DM has on the width and peak of the bursts. Intrinsically narrow pulses with large DM, such as FRBs,  will see a greater increase in signal-to-noise, as well as a reduction in the width, over the directly integrated bandwidth. Figure \ref{fig:DMcorrected} illustrates this. We see that for all of the candidates correcting the DM, $\rm DM_{corr}$, decreases the pulse width and increases the peak, as expected. However, the candidate ``TR" is clearly an outlier. The pulse width is intrinsically much narrower than the others (Table \ref{tab:transients}). 

\begin{figure*}
\includegraphics[width=\textwidth]{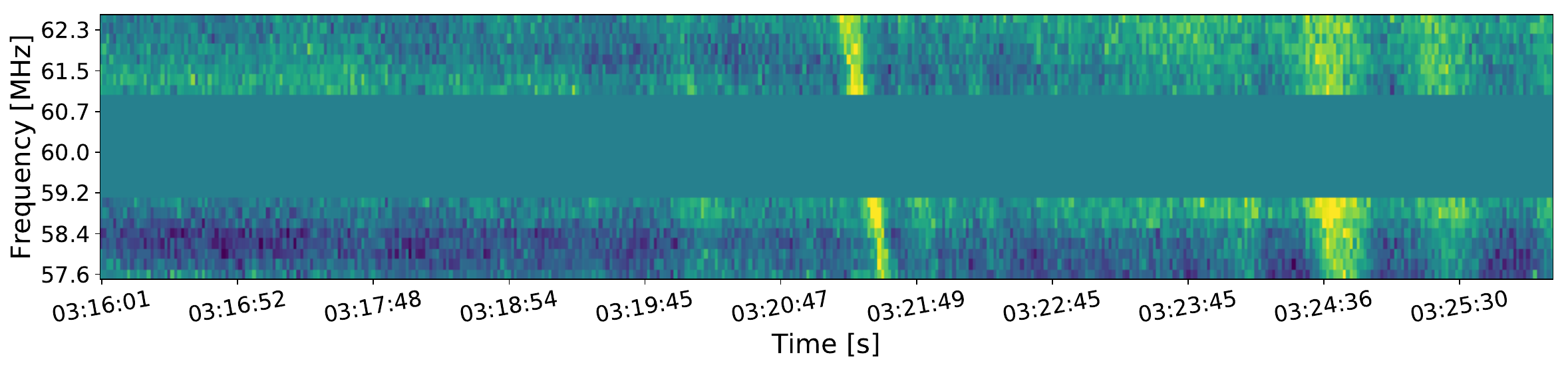}
\caption{Time-frequency plot for the best extragalactic transient candidate detected during our survey (centre). The two sources following 03:24:36 are more consistent with scintels, given thier lack of a well defined dispersion delay,and timescale, though no scintillating source is clearly associated (Fig. \ref{fig:TR_image}). Variation in the candidates flux density peak from 57.8 to 62.5\,MHz is consistent with a flat spectrum after correcting for the time and frequency dependent background variation.}
\label{fig:TRtimefreq}
\end{figure*}

\begin{figure}
\includegraphics[width=\columnwidth]{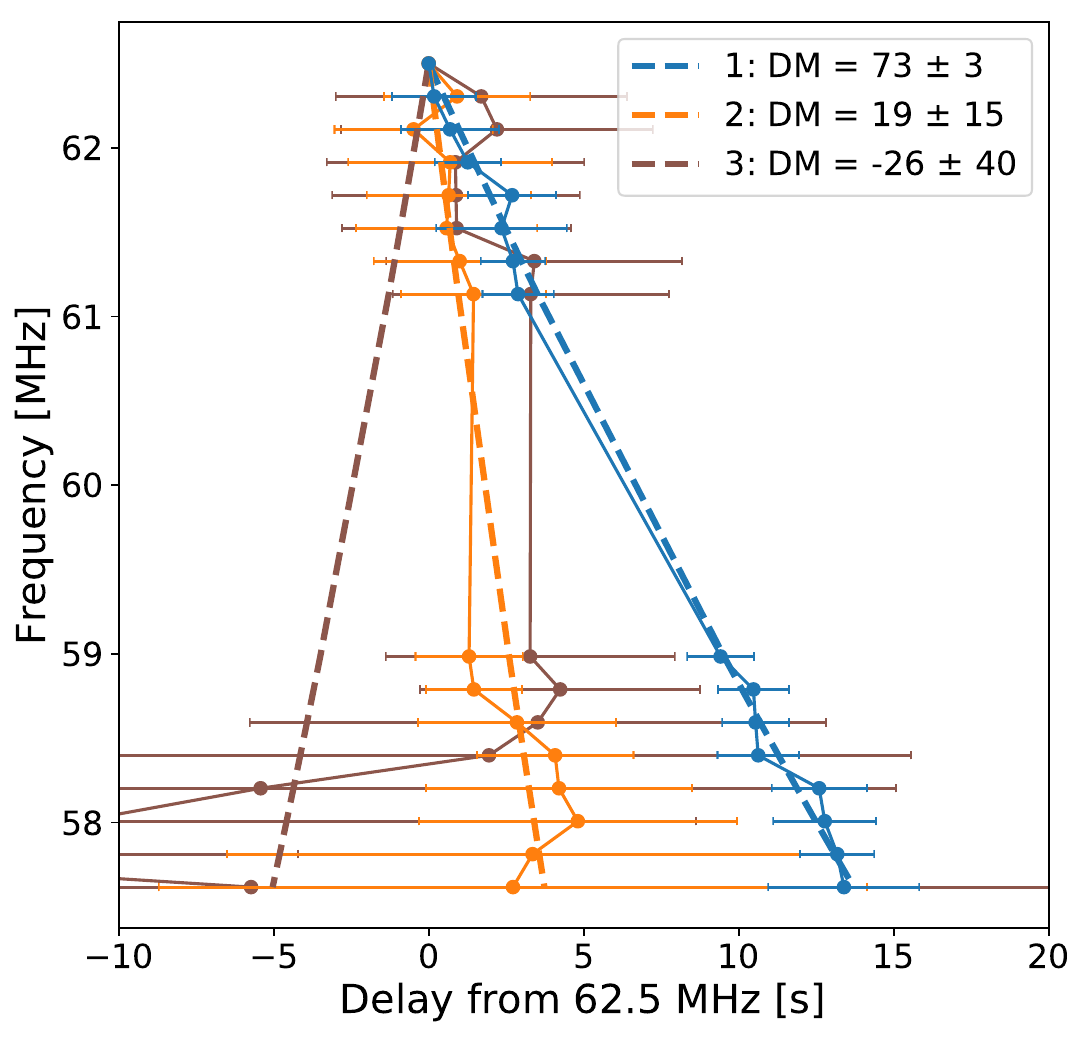}
\caption{The frequency dependent delay in the time of flux density peak for the three sources in Fig. \ref{fig:TRtimefreq}, and the best fit DMs, and numbered in chronological order.Transient candidate 1, shows clear time-frequency dependence consistent with a dispersion delay, whereas candidates 2 and 3 are do not. }
\label{fig:DMdelayTR}
\end{figure}

We illustrate this in more detail with 
the time-frequency plot for this outlier candidate in Fig.~\ref{fig:TRtimefreq}, which also contains additional transient signal after 03:24:36 UTC that is more consistent with the 'usual' ionospheric scintels. Though these appear near in time and space, the are clearly quite different in their characteristics. 
The first one is sharply defined and short in each subband, and the
delay across the full band is larger than the width. The other two are much less sharp, lasting 30--40 seconds, and with a delay across
the full band, if any, that is much less than the width, similar to the singular magnifications. 
In Fig.~\ref{fig:DMdelayTR} we quantify this by showing the result of inferring a DM from time lag between the 16 subbands, measured by cross correlating each subband with the highest frequency, for each candidate in Fig. \ref{fig:TRtimefreq}. 
The first event (blue) has a tightly constrained delay curve, defined fairly sharply with a much lower uncertainty than the following two,
as well as a value of the DM that matches the expectation for an extragalactic source in this sky location (Fig.~\ref{fig:DMcompare}).

\begin{figure}
\includegraphics[width=\columnwidth]{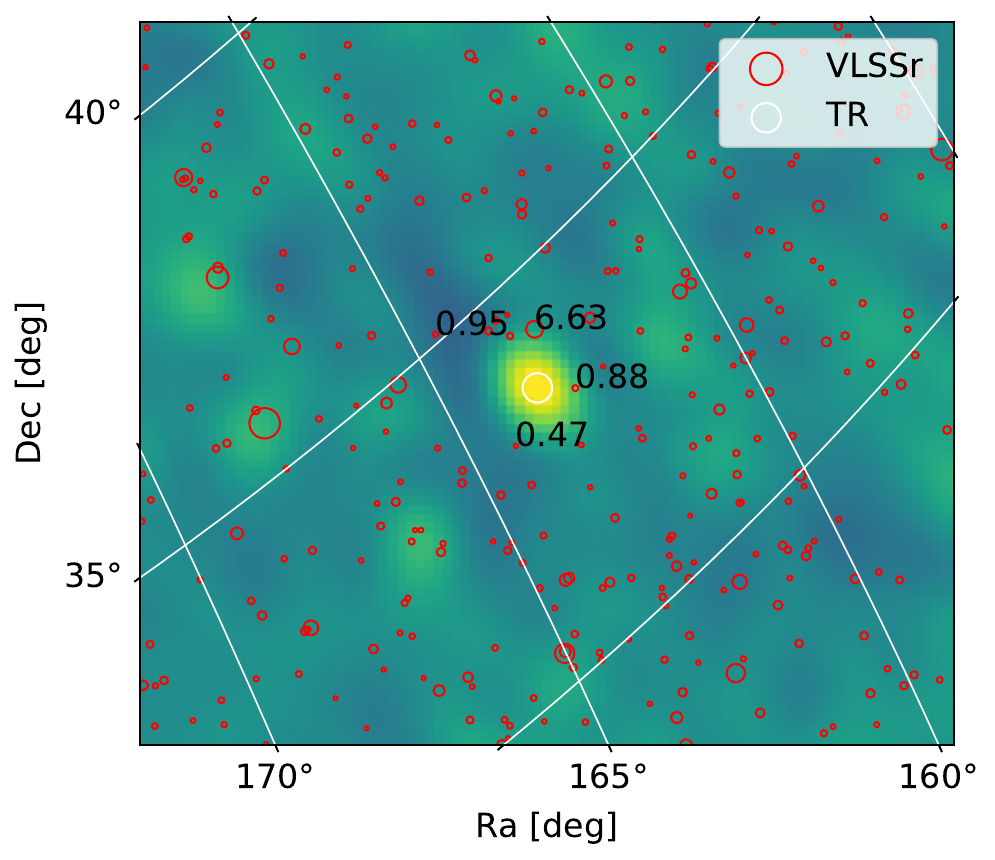}
\caption{The image of the transient candidate at its peak flux density. The white circle radius corresponds to the position uncertainty. The red circles indicate VLSSr sources, with radii are proportional to their flux density. The flux density of all VLSSr sources within a 1 degree radius from the transient candidate have their flux densities labelled.}
\label{fig:TR_image}
\end{figure}

Of course, this explanation implies that the occurrence of two `ordinary' foreground scintels at the same location is a coincidence, and one might reasonably ask whether it is too much of one.
To answer this, we first calculate the sky density of scintels during the time of the transient candidate `TR', and find a chance alignment probability of magnifying plasma lens at the time of this candidate of $2 \times 10^{-3}$ per second. 
During the 10 minutes illustrated in Fig. \ref{fig:TRtimefreq} one would therefore expect $1.3 \pm 1$ scintels, in fine agreement with the two observed. 
However, this agreement is somewhat undermined by a closer look at the image of this candidate (Fig.~\ref{fig:TR_image}): there are no bright steady sources within the PSF of the transient (of the nearest sources within 1 degree, only one is greater than 1\,Jy), so the scintels have to have magnifications of 200 or more. Such scintels
are rare, more commonly the source being magnified is brighter and easily identified in the AARTFAAC data \citep{2021MNRAS.tmp.1149K}.
Unfortunately we do not yet have good enough statistics of bright scintels to quantify this better, but on the whole the coincidence of the bright scintels with the candidate extragalactic source is somewhat uncomfortable. 
Finding a few more examples and/or better characterising the likelihood and magnification distribution of scintillation storms will be needed to resolve this.

\begin{table*}
\begin{center}
\caption{Our sample of high interest transient candidates, defined as bursts with a large and negative inter-frequency delay, potentially corresponding to a physical dispersion delay. The Model DM is the sum of the NE2001 for the Galactic ISM and the halo model. The flux density given is the peak of the light-curve, generated by measuring the integrated flux after dedispersing with the fit DM. Similarly, the width is the FWHM of the lightcurve feature, after DM correction.}
\begin{tabular}{l l l l l l l l}
\hline \hline
Date Time  &  Label  &     Ra    &  Dec  &    $\rm DM_{\rm model}$ &   $\rm DM_{\rm fit}$   & Flux density & Width \\
UTC          &      &     Degrees   &  Degrees  &   $\mathrm{~pc~cm^{-3}}$  &    $\mathrm{~pc~cm^{-3}}$ & Jy & Seconds \\
\hline
2017-02-25 03:21:27    & TR  & 163.37 & 33.73  &    60 &  $ 73  \pm  3 $ &  $80 \pm 30$ & 7.7  \\
2017-02-25 04:19:11    & 2   & 215.16 & 26.75  &    56 & $ 108 \pm  4 $ &     $88 \pm 24$ & 42.6 \\
2017-02-25 04:36:41    & 12  & 236.77 & 17.75  &   71 & $ 56  \pm  7 $ &   $100 \pm 30 $ & 20.8 \\
2018-09-23 04:48:13    & 1   & 36.66  & 26.85  &   82 &  $ 18  \pm  21 $ &  $65 \pm 26 $ & 33.5 \\
2018-09-23 05:29:44    & 5   & 60.98  & 22.69  &    105 &  $ 34  \pm  5 $ &  $62 \pm 14 $ &  23.5 \\
2019-01-01 03:42:16    & 9   & 147.22 & 41.11  &    67 &  $ 108 \pm  5 $ &  $ 70 \pm 30 $ &  32.4 \\
2019-01-01 03:44:08    & 6   & 144.64 & 42.17  &    68  &  $ 60  \pm  4 $ &  $67 \pm 25 $ & 17.5 \\
2019-01-01 03:47:45    & 4   & 157.74 & 34.39  &     62 & $ 99  \pm  3 $ &  $103 \pm 20 $ & 22.1 \\
2019-01-01 05:16:25    & 10  & 173.51 & 10.01  &   350  &  $ 65  \pm  6 $ &   $75 \pm 32 $ & 29.0 \\
2019-02-03 01:22:58    & 8   & 186.33 & 25.90  &    154 & $ 53  \pm  6 $ &   $63 \pm 24 $  & 51.8 \\
\hline \hline
\label{tab:transients}
\end{tabular}
\end{center}
\end{table*}

\section{Discussion}
\label{sec:discussion}

Assuming our strong candidate, illustrated in Fig. \ref{fig:TR_image},  is a true extragalactic flare, we can search for potential associations within the position and distance uncertainty. With a best-fit DM of $73 \pm 3 \mathrm{~pc~cm^{-3}}$, this DM would be extragalactic: in the direction of the source (RA $163.7 \pm 0.5$ deg, DEC $33.8 \pm 0.5$ deg), the total Galactic DM is $30 \mathrm{~pc~cm^{-3}}$ \mbox{\citep{2002astro.ph..7156C}}, and the halo contribution is also $30 \mathrm{~pc~cm^{-3}}$ \citep{2020ApJ...888..105Y}. 
This places the source origin definitely outside the disc of our Galaxy and likely outside of the halo (but, given local variability of sight-lines relative to the DM models, not with very great certainty). Using the DM-distance relation for the IGM, $\mathrm{DM} = 900z \mathrm{~pc~cm^{-3}}$ \citep{2014ApJ...780L..33M, 2019ApJ...885L..24C} gives a redshift value of 0.015, or distance of 63 Mpc \citep{2006PASP..118.1711W}. Given that the foreground DM value is uncertain and that there might be also a contribution from the environment or host of a putative extragalactic event, we should regard this value as indicative of the upper limit to the distance rather than a measured value.

In the Advanced Detector Era (GLADE, v2.3) catalogue of nearby galaxies \citep{2018MNRAS.479.2374D}, there are three potential galaxies within the position uncertainty and distance limit: PGC2039259 at 26.4\,Mpc, NGC3442 at 27.96\,Mpc, and PGC032706 at 55.83\,Mpc, all with flux densities of 0.5\,Jy or less. None of these is detected in our AARTFAAC data, so if the flare originates from one of them, its radio luminosity is more than 200 times that of the quiescent galaxy's total value. Since we cannot identify the event with a quiescent source, the cause of this flare remains rather speculative at the moment. To set the scale, at a distance of 50\,Mpc this flare has a peak radio luminosity (assuming a spectral range of 100\,MHz and isotropic emission) of $3\times10^{37}$\,erg/s and emitted energy only in the radio band of $3\times10^{38}$\,erg. 
By the usual argument of how fast a radio flare can rise from a source of a given size \citep[e.g.,][]{2015MNRAS.446.3687P}, this implies a brightness temperature in excess of $10^{34}$\,K. Even if we do assume the source is on the outskirts of the Galaxy, at 50\,kpc, the implied brightness temperature would still be $10^{28}$\,K. In either case, the emission mechanism underlying this phenomenon then needs to be coherent. This also means the radiation is very likely beamed and thus the stated luminosities and energies are upper limits.

Given these rather strong implications, we prefer to regard this event as a candidate extra-Galactic event for now, and await the discovery of a few more confirming cases and better understanding of the outliers in the frequency-time delay distribution of the magnification events.

Our survey places constraints at shorter timescales and with a very large survey volume, to a depth that is however less than most other surveys. 
Our results are broadly consistent with other low-frequency surveys, which report mostly upper limits to the rate of radio transients below 200\,MHz, at timescales below one minute, \citep{2015JAI.....450004O, 2016MNRAS.458.3506R, 2019ApJ...886..123A}. 
These three works, together with ours, place similar limits on the brightness of a hypothetical population of isotropically distributed, standard candle, transient events. But we shall now compare them in more detail for consistency also with some detections in this frequency range, by taking these differences into account.

We have observed a single example of a potential transient, with a flux density of
$80\pm 30$\,Jy and lasting 7.7\,s, and a dispersion delay across our frequency range indicative of an extragalactic origin. 
If this transient is a member of a population of sources that is isotropic on the sky and uniform in space density, the observed power-law brightness distribution $(\log(N>S)\propto S^{-p})$ of such a population would have a so-called Euclidean index $p=3/2$. 
That theoretical brightness distribution is plotted as the red dash-dotted line in the righthand panel of  Fig. \ref{fig:sensitivity}. 
One previous upper limit lies below this line, from the 28\,s-timescale MWA survey by \cite{2016MNRAS.458.3506R}, at 182\,MHz. However, to see whether this indicates a flatter than
Euclidean brightness distribution we must account for the fact that our transient is shorter than the 28\,s cadence of the MWA survey. The fluence of our transient was $580\pm 150$\,Jy\,s, and so its flux averaged
over a 28\,s time bin was $20\pm 6$\,Jy. If we use that value and $p=3/2$ to extrapolate to the sensitivity threshold of 0.285\,Jy of the MWA survey, we predict a surface density of $2.4\times10^{-7}$\,deg$^{-2}$, somewhat below their upper limit of $4.1\times10^{-7}$\,deg$^{-2}$. Their upper limit is thus consistent with our detection. (Note we have taken the optimistic view that the event would fall entirely within one MWA time bin; more realistically it would be split over two and this would further reduce the detection probability, see \cite{2016MNRAS.459.3161C}.) The MWA survey was at a three times higher frequency than ours, so this marginal consistency implies that if our source is part of Euclidean flux distribution, it cannot have a strongly rising, self-absorbed, spectral index between 60 and 182\,MHz.

The population of extreme-fluence pulses from PSR\,B0950$+$08 \citep{2020MNRAS.497..846K} 
is of course entirely from one source, and so the rate or surface density averaged over the sky should be taken with a grain of salt. Also, this pulsar is very nearby (DM$=3.0$\,pc\,cm$^{-3}$) and thus is not dispersed enough for the pulse to be smeared over multiple time bins across our band. Most other known giant-pulse emitting pulsars have rather higher DM, and we are now searching our data with de-disperion to their known DMs in order to see whether this phenomenon extends to other pulsars.

The population of isolated magnifications of background sources, due to turbulent ionized plasma in the ionosphere-plasmasphere region \citep{2021MNRAS.tmp.1149K}, has an
effective surface density that agrees well with the non-detection by \cite{2019ApJ...886..123A}, given the rate and timescale probed. 
Many characteristics of the cosmic transient detected by LWA1 and LWA-SV \citep{2019ApJ...874..151V}, including timescale, burst shape, and brightness, fit this population well.
Although the rate is somewhat larger than our distribution would predict, this could be consistent with the fact that a plasma lensing is stronger at lower frequencies \citep[e.g.,][]{1998ApJ...496..253C}, and the LWA1/LWA-SV survey was conducted at 34\,MHz.
The non-detections of this particular phenomenon by both \cite{2015JAI.....450004O} and \cite{2019ApJ...874..151V} are consistent with the distribution we see. While the survey by \cite{2015JAI.....450004O} probes a very large volume at 60\,MHz, it did not have the sensitivity required. 
And \mbox{\cite{2019ApJ...874..151V}}, which was much more sensitive at the same frequency, searched insufficient volume, for the timescale analyzed.

\cite{2016MNRAS.456.2321S} detected a single transient event with a flux density of 15--25\,Jy in a survey that used 2149 snapshots of 179 square degrees each around the North celestial Pole, each lasting 11\,min. 
Given the positive detection and surveyed volume, the rate was determined to be $3.9^{+14.7}_{-3.7}\times10^{-4} ~\mathrm{day}^{-1}\mathrm{deg}^{-2}$.
If we translate this rate to our typical sensitivity limit of 60\,Jy with a Euclidean source count distribution, $N(>S) \propto S^{-3/2}$, we expect about 8 similar  events in the volume of our survey. This means that our non-detection has quite a low probability, $2.8\times10^{-4}$, under the Euclidean assumption. 
Since the frequencies of the two surveys are the same, spectral index cannot be a confusing issue here and so we conclude that between 15 and 60\,Jy, the source counts for this population must be quite a bit steeper than Euclidean, $p\leq -2.5$. 
Our next survey, which uses an additional six LOFAR stations and has a ten times better sensitivity (light blue dashed curve in Fig.~\ref{fig:sensitivity}) does go down to below the flux level of the \cite{2016MNRAS.456.2321S} transient and should therefore detect this population.

The detection of FRBs at the lowest frequencies remains an important goal for constraining the rate and spectral shapes of these events, and thereby perhaps their origin and emission mechanism.
Therefore we evaluate to what degree our survey is sensitive to the brightest tail end of the FRB population.
While FRBs have been detected between 110-188\,MHz with LOFAR \citep{2021ApJ...911L...3P}, these originated only from FRB 20180916B, a single repeating FRB.
We therefore use the all-sky FRB rate measured by CHIME \citep{2019Natur.566..230C}, who have reported detections of many single and repeating FRBs across the Northern Hemisphere detected at frequencies down to 400\,MHz,  the bottom of the CHIME bandpass.
The all-sky rate they give is $300~\mathrm{day}^{-1}$ above a flux density limit of 1\,Jy, consistent with the value of $400~\mathrm{day}^{-1}$ by \cite{2016MNRAS.460.1054C}. 
The observing frequencies are too different to neglect, but since FRB spectra have a wide range of measured slopes \citep[][and references therein]{2019A&ARv..27....4P}, both rising and  falling to higher frequencies, we cannot do much better than assume a flat spectrum. 
Next, we must scale the sensitivity threshold for the burst peak flux density to 60\,kJy. 
This results from our 1\,s integration, with an average 1\,ms intrinsic burst width. 
On the bright end of the distribution, the FRB rate seems to follow $ R(> F) \propto F^{-2.2}$ \citep{2019A&ARv..27....4P}.
Therefore, the expected rate of FRBs greater than 60\,kJy is $9\times10^{-9}~\mathrm{day}^{-1}\mathrm{sky}^{-1}$, which translates to a surface density per snapshot of $2.6\times10^{-18}~\mathrm{deg}^{-2}$; this yields a completely negligible  $2\times10^{-8}$ expected FRB detections in our survey. It is therefore not surprising that we did not detect any FRBs, even if we optimistically assumed a fraction of them to have steep spectral indices, especially since we also neglected the deteriorating effect of dispersion smearing.

\section{Conclusions}
\label{sec:conclusion}

We have presented results of the first survey with the AARTFAAC instrument for rare, bright, and fast radio transients at 60\,MHz and shown for the first time two significant-sized populations of such radio transients: extreme pulses from the nearby pulsar B0950$+$08 and strong, short-duration magnifications of known radio sources caused by space weather, as well as a single candidate extragalactic transient, possibly the first of another  class of transient. 
We set a strong upper limit of about 1.1/sky/day to any yet undiscovered types of transient with peak fluxes above 60\,Jy and durations of seconds to minutes.
We also discuss the power of our survey to constrain the bright end of the FRB population, and find that this is very little. But in the case of the enigmatic 5\,min transient found by \cite{2016MNRAS.456.2321S}, we can constrain the slope of the brightness distribution to be much steeper than Euclidean because we did not detect any such transients. 
However, since by far most false positives in our search come from detections in only one band, because RFI is typically narrow band (see Table \ref{tab:Filterlist}), we have to reject all narrow-band sources. Therefore, if these transients are also narrow-band, they could only be constrained from a site with much lower RFI or a search with narrower frequency channels.
Finally, we set very low upper limits to the surface density and rate of any other, yet unknown types of low-frequency radio transient.

We also find one case of a 7.7\,s, 80\,Jy flare that shows all the hallmarks of being dispersed, with a DM that likely puts it at extragalactic distance, or at the very least in the outskirts of our own Galaxy. We find no underlying radio source brighter than 0.5\,Jy. We withhold final judgement on whether it really is dispersed until we have collected more events, since there are some outliers in the distribution of magnification events that have similar time delays to this event. If it is indeed a distant source, its emission is quite extreme, requiring a brightness temperature of 10$^{28-34}$\,K.

There are still quite a number of steps of improvement possible: first, we have about as much 6-station AARTFAAC data yet to be analysed as we report here, and we can search the data with crude de-dispersion in order to look for pulses like those of B0950 in other pulsars. Furthermore, we have calibrated the new 12-station AARTFAAC and performed a first short survey with it \citep{2021arXiv210315160S}, and are about to embark on a much longer survey with this roughly ten times more sensitive array. This should, for example, allow us to detect a population of sources like the one found by \cite{2016MNRAS.456.2321S} and expand on, or disprove, the existence of fast extragalactic transients like our current candidate; and perhaps it will unveil more unknown unknowns.

\nocite{2016MNRAS.456.2321S}
\nocite{2014MNRAS.438..352B}
\nocite{2016MNRAS.459.3161C}
\nocite{2011MNRAS.412..634B}
\nocite{2010AJ....140.1995L}
\nocite{2009ApJ...696..280H}

\section*{Data Availability}

The data underlying this article can be shared on reasonable request to the corresponding author. Additionally it some data products are available through the Zenodo open-acces data repository DOI:10.5281/zenodo.4778360.

\section*{Acknowledgements}

AARTFAAC development and construction was funded by the ERC under the Advanced Investigator grant no. 247295 awarded to Prof. Ralph Wijers, University of Amsterdam; This work was funded by the Netherlands Organisation for Scientific Research under grant no. 184.033.109. We thank The Netherlands Institute  for Radio Astronomy (ASTRON) for support provided in carrying out the commissioning observations. AARTFAAC is maintained and operated jointly by ASTRON and the University of Amsterdam.

We would also like to thank the LOFAR science support for their assistance in obtaining and processing the data used in this work. We use data obtained from LOFAR, the Low Frequency Array designed and constructed by ASTRON, which has facilities in several countries, that are owned by various parties (each with their own funding sources), and that are collectively operated by the International LOFAR Telescope (ILT) foundation under a joint scientific policy.

This research made use of Astropy,\footnote{http://www.astropy.org} a community-developed core Python package for Astronomy \citep{astropy:2013} \citep{astropy:2018}, as well as the following  KERN \citep{molenaar2018kern}, Pandas \citep{mckinney-proc-scipy-2010}, NumPy \citep{2011CSE....13b..22V}, and SciPy \citep{citescipy}. Accordingly, we would like to thank the scientific software development community, without whom this work would not be possible. 




\bibliographystyle{mnras}
\bibliography{survey_paper}


\label{lastpage}
\end{document}